\documentclass[11pt,a4paper]{article}

\usepackage[english]{babel}

\usepackage[T1]{fontenc}
\usepackage[math]{iwona} % Mittelbach (2023, 12.5.5)

\usepackage{amsmath,amssymb}

\usepackage{graphicx}

\usepackage{placeins}

\usepackage{natbib} % Mittelbach (2023, 16.4)
\usepackage{IEEEtrantools}

\usepackage{csquotes}

\usepackage{tikz} \usetikzlibrary{shapes}

\usepackage{hyperref} % Oetiker (2023, 5.3)

\usepackage{calc}

\usepackage[paper=a4paper,portrait, tmargin=30mm, bmargin=30mm, rmargin=40mm, lmargin=25mm]{geometry} % Mittelbach (2023, 5.2.4)

\usepackage{tabls}

\newlength\micolumna
\newcommand{\figwidth}{0.6\columnwidth}

\title{Economics of Integrated Sensing and Communication service provision in 6G networks\\

\thanks{This work was supported through Grant PGC2018-094151-B-I00 and PID2021-123168NB-I00, funded by MCIN/AEI, Spain/10.13039/ 501100011033 and the European Union A way of making Europe/ERDF, and Grant TED2021-131387B-I00, funded by MCIN/AEI, Spain/ 10.13039/501100011033 and the European Union NextGenerationEU/ RTRP.}
}

\author{Luis Guijarro\textsuperscript{\textdagger}, Maurizio Naldi\textsuperscript{\textdaggerdbl}, Vicent Pla\textsuperscript{\textdagger}, José-Ramón Vidal\textsuperscript{\textdagger}\\
\textsuperscript{\textdagger} Universitat Politècnica de València, \textsuperscript{\textdaggerdbl}LUMSA}

\date{May 16, 2024}

\begin{document}

\thispagestyle{empty}

\begin{tabular*}{\textwidth}[t]{|p{\micolumna}|p{\micolumna}|}
	\hline
	\includegraphics[width=\micolumna,trim=2cm 17cm 2cm 2cm,clip]{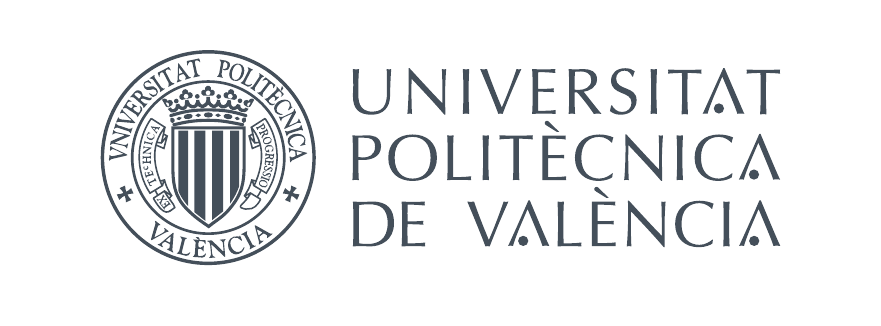}& 
	\raisebox{1.6cm}{\parbox{\micolumna}{Departamento de Comunicaciones}}\\
	\hline
	\multicolumn{2}{|c|}{}\\
	[0.20\textheight]
	\multicolumn{2}{|p{\textwidth-\tabcolsep*2}|}
	{\huge \begin{center}
Economics of Integrated Sensing and Communication service provision in 6G networks
\end{center}}\\
	\multicolumn{2}{|c|}{\huge (Document NETECON132D)}\\
	[0.20\textheight]

	\hline
	Authors: Luis~Guijarro,  & Outreach: Public\\ 
	\cline{2-2}
	Maurizio Naldi (LUMSA, Italy)  & Date: May 16, 2024 \\ 
	\cline{2-2}
	 Vicent Pla, José Ramón Vidal & Version: c\\
	\hline
	\multicolumn{2}{|p{\textwidth-\tabcolsep*2}|}{Copyright notice: \enquote{\copyright 2024 IEEE.  Personal use of this material is permitted.  Permission from IEEE must be obtained for all other uses, in any current or future media, including reprinting/republishing this material for advertising or promotional purposes, creating new collective works, for resale or redistribution to servers or lists, or reuse of any copyrighted component of this work in other works.}}\\
	\hline
\end{tabular*}

\clearpage

\setcounter{page}{1}

\maketitle

\begin{abstract}
In Beyond5G and 6G networks, a common theme is that sensing will play a more significant role than ever before. Over this trend, Integrated Sensing and Communications (ISAC) is focused on unifying the sensing functionalities and the communications ones and to pursue direct tradeoffs between them as well as mutual performance gains. 
We frame the resource tradeoff between the S\&C functionalities within an economic setting.
We model a service provision by one operator to the users, the utility of which is derived from both S\&C functionalities. The tradeoff between the resources that the operator assigns to the S\&C functionalities is analyzed from the point of view of the service prices, quantities and profits.
We demonstrate that equilibrium quantities and prices exist. And we provide relevant recommendations for enforcing regulatory limits of both power and bandwidth. 
\end{abstract}

\section{Introduction}

Next-Generation wireless networks (such a beyond 5G (B5G) and 6G) have been envisioned as key enablers for many emerging applications. Among many visionary assumptions about these networks, a common theme is that sensing will play a more significant role than ever before. Indeed, the technological trends clearly show that we are ready to embrace this new sensing capability in the forthcoming B5G and 6G eras. Both radio sensing and communication systems are becoming increasingly similar in terms of hardware architectures and signal processing. This offers an exciting opportunity for implementing sensing by utilizing wireless infrastructures, in such a way that future networks will go beyond classical communication and provide ubiquitous sensing services to measure or even to image surrounding environments.

Over the above trend, Integrated Sensing and Communications (ISAC) is focused on unifying the sensing operations and the communications ones and to pursue direct trade-offs between them as well as mutual performance gains~\citep[Chap. 21]{tong2021}. ISAC will offer advantages in several case studies, such as sensing as a service, smart home and in-cabin sensing, vehicle to everything, smart manufacturing, geoscience, environmental sensing and human-computer interaction~\citep{liu2022}.

One fundamental design issue in ISAC emerges when wireless resources are shared between the sensing and the communication (S\&C) functionalities, since tradeoffs between, often contradictory, S\&C objectives and metrics should be solved. This is a tradeoff raised at the physical level.

The present work frames the physical tradeoff between the S\&C functionalities within an economic setting, which will contribute to fill a research gap that the authors have identified, as detailed below. We model a service provision by an operator to the users, the utility of which is derived from both S\&C functionalities. The tradeoff between the resources that the operator assigns to the S\&C functionalities is analyzed from the point of view of the service prices, quantities and profits.

The structure of the manuscript is as follows. Section~\ref{sec:related} describes the related work. Section~\ref{sec:model} presents the system and economic models and states the analysis. Section~\ref{sec:results} presents and discusses the results, and Section~\ref{sec:conclusions} provides the conclusions.

%--------------------------------------------------------------
%--------------------------------------------------------------
\section{Related work}\label{sec:related}

In this section, we provide a brief review of the literature of interest, considering the following three streams: the integration of communications and sensing (radar) functions in the 6G infrastructure; the economic analysis of ICT systems; the economic analysis of passive/bistatic radars.

The integration of sensing and communications functions (aka as ISAC) is a major feature of the advent of the 6G infrastructure \citep{wei2023integrated,kim2022survey}. The feasibility of using 6G frequencies for passive radars has been examined in \citet{lingadevaru2022feasibility}. The co-habitation of different functions in the same network has posed several resource sharing problems. For example, a time-sharing discipline has been proposed and optimized in \citet{xie2023optimal}. Signal design has been optimized in \citet{wu2023joint}. The optimal allocation of the overall transmitting power between the sensing and the communication tasks has been investigated in \citet{liu2022}, while the capacity-distortion trade-off in a memoryless channel has been evaluated in \citet{kobayashi2018joint}. In all these works, optimization is achieved just under a technical performance viewpoint with no economic considerations. A high-level analysis of the 6G network infrastructure and its cost drivers (coverage and antennas, backhaul, spectrum and edge computing units) is conducted in \citet{kokkinis20236g} for some use cases, where sensing is, however, not considered.

On the other hand, the economic analysis of communications services has a long history (see, e.g., the most recent papers \citet{guijarro2021,flamini2023optimal} and the textbooks \citet{courcoubetis2003,maille2014telecommunication}).

There is not an equivalent history in radar systems, which have traditionally been designed considering just the technical dimension. For example, the reference handbook by Skolnik does not mention the words \textit{cost} or \textit{price} \citep{skolnik2015radar}. However, the recent resurgence of passive/bistatic radar must rely on some economic advantage, like, e.g., the exploitation of transmitters of opportunity \citep{willis2007advances}. Low cost is mentioned as one of the most interesting features that would push the replacement of traditional active radars by passive radars \citep{judice2023current}. Examples of economics-aware approaches to the design of passive radar systems are shown in \citet{chang2016fault,wang2015minimum,xu2019minimum}, where the optimal placement of passive radars to achieve belt barrier coverage is sought after; however, the cost of the radar systems involved is considered fixed and is not a decision variable. A similar optimization task, where the placement of fixed cost devices is considered for WiFi-based passive bistatic radars, is investigated in \citet{ivashko2014receivers}. In none of these works, the optimization of the radar's characteristics (e.g., the transmitting power) is considered.

There has been no attempt, as far as the authors are aware, to address the economics of the combined sensing and communication-related aspects of an ISAC-based service provision. This manuscript is a first attempt to frame the ISAC-based service provision in an economic setting, that is, one where both the communication and sensing resources procurement and the service provision are mediated by prices.

%--------------------------------------------------------------
%--------------------------------------------------------------
\section{Model}\label{sec:model}

We model an operator that uses power and bandwidth resources for providing a service supported by S\&C functionalities. The operator is a monopolist in the service market, but not in the input-factor market, where it acquires its resources at given prices. 

\subsection{System model}

We model a joint communication and passive radar system as the support of the ISAC service offering, based on~\citet{chalise2017b} and~\citet{hack2013}.

Passive radar systems, compared to conventional active radar systems, which typically operate in a mono-static mode and emit strong signals with a wide signal bandwidth, use broadcast signals that in general are very weak and have an extremely narrow bandwidth.

The ISAC transmitter emits a sensing waveform to detect targets using a portion of its total power budget, and emits a communication waveform using another portion. For simplicity, we assume in this work that the two signals are scheduled over orthogonal frequency resources such that they do not interfere with each other. The ISAC sensing receiver receives direct and target sensing signals through the direct and the surveillance paths, respectively, and wishes to detect the presence of a target in the latter; on the other hand, the communication receiver receives a communication signal, which contains useful information.

The communication metric depends on the achievable bit rate. According to the Shannon-Hartley theorem, this bit rate (in bps) is given by:
\begin{equation}
R_c  =  W_c \log_2 \left( 1 + \frac{S}{N} \right).
\end{equation}
%R_c  =  W_c \log_2 \left( 1 + \frac{E_b}{N_0} \right).
%
% We assume that a bandwidth $W_c$ is needed in order to transmit a symbol rate $\frac{1}{W_c}$, which is a usual and more conservative choice than the Nyquist rate. 
%
Where $S$ is the average received signal power and $N$ the average power of the noise, both over the bandwidth $W_c$. The power at the transmitter is  $P_c$, and the communication channel gain is $\gamma_c$, the bit rate is
\begin{align}\label{eq:Rc}
R_c & = W_c \log_2 \left( 1 + \frac{P_c \gamma_c^2 }{N_0 W_c} \right) \nonumber \\
& = W_c \log_2 \left( 1 + \frac{P_c \tilde{\gamma_C}}{W_c}  \right),
\end{align}
where $\tilde{\gamma}_C \equiv \frac{\gamma_c^2}{N_0}$.

The sensing metric is the radar detection probability. If a power $P_r$ is used at the transmitter, the surveillance channel gain is $\gamma_t$, the sensing signal bandwidth is $W_r$, and we assume that the direct-path signal-to-noise-ratio is high enough, the detection probability is given by:
\begin{equation}
P_D = Q_1 \left( \sqrt{2 \frac{P_r \gamma_t^2}{N_0 W_r} } , \sqrt{2 \gamma} \right),
\end{equation}
where $Q_1(a,b)$ denotes the first-order Marcum Q-function with parameters $a$ and $b$, and $\gamma = - \log P_{FA}$, and $P_{FA}$ is the probability of false alarm, which is taken as a parameter.
We assume that a bandwidth $W_r$ is needed in order to transmit a symbol rate $W_r$, 
%which is a usual and more conservative choice than the Nyquist (signaling) rate $2 W_r$, 
that the detection takes into account a time period $T$, 
and that $P_r$ is proportional to the number of symbols. Thus, $P_r$ is proportional to $W_r T$, and $P_D$ does not depend on $W_r$. Therefore, the only requirement is that $W_r$ be a feasible design choice; it is taken as a parameter to be included in $\gamma_T \equiv \frac{\gamma_t^2}{N_0 W_r}$, and the probability of detection is
\begin{equation}\label{eq:PD}
P_D = Q_1 \left( \sqrt{2 P_r \gamma_T} , \sqrt{2 \gamma} \right),
\end{equation}

\subsection{Economic model}

We envision a scenario where an operator provides the ISAC service to a population of users. 

The service provision is modeled as the selling of two commodities~\citep[Chap.~3]{courcoubetis2003}: the radar transmission power ($P_r$) and the communication bit rate ($R_c$), which are priced at unit prices $p_1$ and $p_2$, respectively. The first commodity is delivered directly  by the operator without any transformation, while the second commodity, based on~\eqref{eq:Rc}, is ``produced" from two ``input factors": the communication transmission power $P_c$ and the bandwidth $W_c$.

We, first, derived the expression for the service demand by the users, under the assumption of utility maximization, and then derive the expression for the service supply by the operator, under the assumption of profit maximization.

\subsubsection{Users' demand}

We borrow a representative consumer model for the users of the ISAC service~\citep[Chap.~4]{mas1995}, with a quasilinear utility that combines linearly a sensing metric $\theta$, which depends on the first commodity, and a communication metric $\eta$, which depends of the second commodity. More specifically, the representative user (hereafter, the user) has a utility:
\begin{equation}\label{eq:utility}
U = \alpha \theta ( P_r ) + \beta \eta ( R_c ) - p_1 P_r - p_2 R_c.
\end{equation}
The sensing metric $\theta$ is the probability of detection~\eqref{eq:PD}:
\begin{equation}\label{eq:sensing utility}
\theta ( P_r ) = Q_1 \left( \sqrt{2 P_r \gamma_T} , \sqrt{2 \gamma} \right).
\end{equation}
And the communication metric $\eta$ has a logarithmic dependence on the bit rate $R_c$~\citep{reichl2013}:
\begin{equation}\label{eq:commun utility}
\eta ( R_c ) = \log \left( 1 + R_c \right) 
\end{equation}

\begin{figure}[t]
\begin{center}
\includegraphics[width=\figwidth]{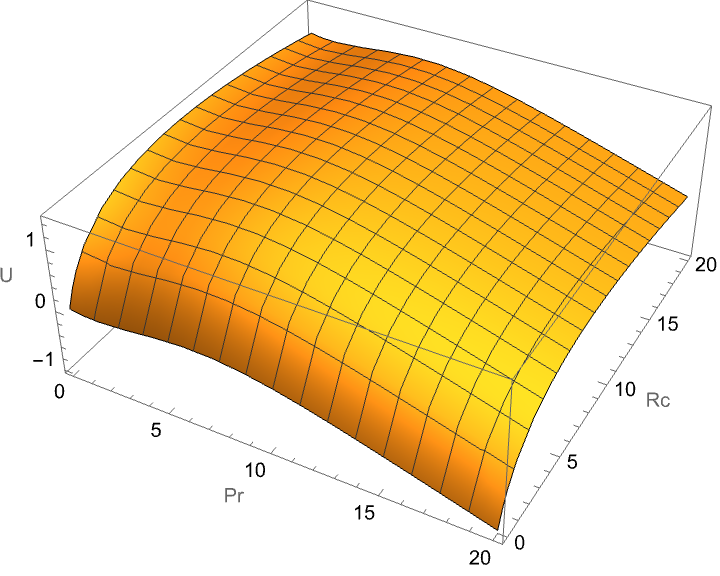}
\caption{Utility as a function of $Pr$ and $R_c$, for $p_1=0.1$ and $p_2=0.1$}\label{fig:utility}
\end{center}
\end{figure}

The utility is plotted in Fig.~\ref{fig:utility} as a function of the commodities $P_r$ and $R_c$, where parameters are set to the default values of Section~\ref{sec:results}.

We assume that the user takes prices $p_1$ and $p_2$ as given, and therefore the decision on how much of commodities $P_r$ and $R_c$ to demand for given prices is the solution of the following utility maximization problem:
\begin{IEEEeqnarray} {rCCl}\label{eq:UMP}
 \underset{P_r,R_c}{\max} & & U \\
 \text{subject to } & &  P_r \geq 0 , R_c \geq 0. \nonumber
\end{IEEEeqnarray} 
Since the utility expression~\eqref{eq:utility} is separable in $P_r$ and $R_c$, two independent utility maximization problems can be stated:
\begin{IEEEeqnarray} {rCCl}\label{eq:UMP1}
 \underset{P_r}{\max} & & \qquad \alpha \theta ( P_r ) - p_1 P_r \\
 \text{subject to } & &  P_r \geq 0, \nonumber
\end{IEEEeqnarray} 
and
\begin{IEEEeqnarray} {rCCl}\label{eq:UMP2}
 \underset{R_c}{\max} & & \qquad \beta \eta ( R_c ) - p_2 R_c. \\
 \text{subject to } & &  R_c \geq 0. \nonumber
\end{IEEEeqnarray} 

The solution to each utility maximization problem yields a demand function, which relates price and optimum quantity. We are interested in interior maxima, so that we seek solutions to the unconstrained first-order conditions (FOCs). 

As regards problem~\eqref{eq:UMP2}, the FOC gives
\begin{equation}\label{eq:demand2}
p_2 = \frac{\beta}{1+R_c}
\end{equation}
which gives a global maximum $R_c^*$, since the objective function is concave.

As regards problem~\eqref{eq:UMP1}, the FOC gives
\begin{equation}\label{eq:demand1}
p_1 = -\alpha \gamma_T Q_1 \left( \sqrt{2 P_r \gamma_T} , 2 \gamma \right) + \alpha \gamma_T Q_2 \left( \sqrt{2 P_r \gamma_T} , 2 \gamma \right),
\end{equation}
which gives a solution $P_r^*$ that is a global maximum only for $P_r^*$ greater than a threshold value. Below the threshold, the solution is only a local maximum and $P_r^*=0$ is the global maximum, instead. However, $P_r^*=0$ is not valid, since the sensing metric has been obtained under the assumption of high signal-to-noise-ration at the sensing receiver. We hereafter assume (and check a posteriori) that~\eqref{eq:demand1} gives a valid solution.

Expressions~\eqref{eq:demand1} and~\eqref{eq:demand2}, where price is the dependent variable, are properly called inverse demand function. We will use them below and they will be denoted $p_1(P_r)$ and $p_2(R_c)$.

\subsubsection{Service provider's decisions}

The operator obtain revenues $R$ equal to
\begin{equation}\label{eq:revenues}
R(P_r,R_c) = p_1(P_r) P_r + p_2(R_c) R_c.
\end{equation}

We assume that the operator obtains the commodity $P_r$ directly without any transformation as an input factor at unit price $w_p$.
As regards the commodity $R_c$, we borrow expression~\eqref{eq:Rc}, which relates $R_c$ to $P_c$ and $W_c$, as a production function, where input factors $P_c$ and $W_c$ produce output product $R_c$, that is:
\begin{equation}\label{eq:production}
R_c = \rho (P_c, W_c) =  W_c \log_2 \left( 1 + \frac{Pc \tilde{\gamma}_C}{W_c}  \right).
\end{equation}
Input factor $P_c$ can be obtained also at unit price $w_p$, and $W_c$ at unit price $w_w$. 

Costs $c$ incurred by the operator when procuring itself with input factors $P_r$, $P_c$ and $W_c$ are then given by:
\begin{equation}\label{eq:costs}
c ( P_r, P_c, W_c) = w_p (P_r + P_c) + w_w W_c
\end{equation}

The profits $\Pi$, which are the revenues minus the costs, can finally be expressed as a function of the input factor quantities as follows:
\begin{multline}\label{eq:profits}
\Pi (P_r,P_c,W_c)= p_1 (P_r) P_r + p_2 (\rho (P_c, W_c) ) \rho (P_c, W_c)\\
 - w_p (P_r + P_c) - w_w W_c
\end{multline}

\begin{figure}[t]
\begin{center}
\includegraphics[width=\figwidth]{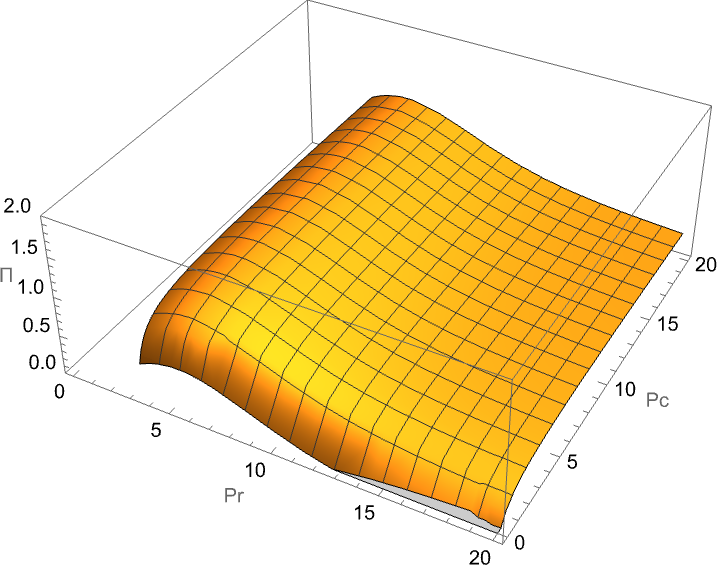}
\caption{Profit as a function of $Pr$ and $P_c$, for $W_c=1$}\label{fig:profits_PrPc}
\end{center}
\end{figure}

\begin{figure}%[t]
\begin{center}
\includegraphics[width=\figwidth]{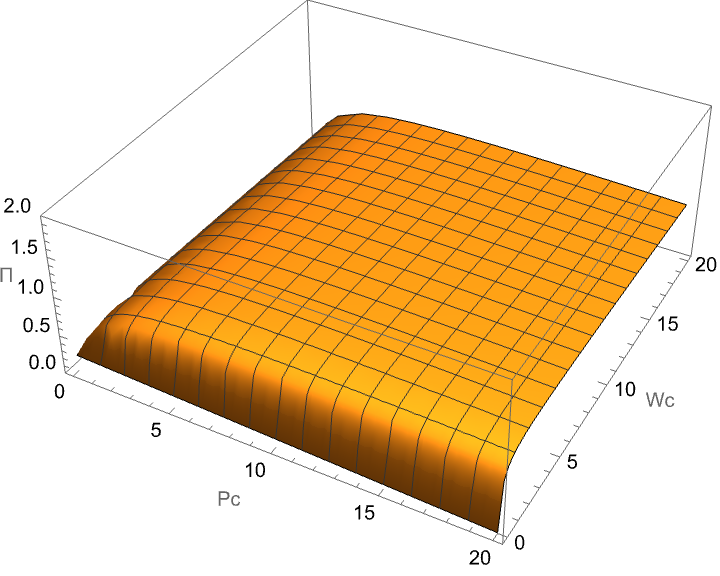}
\caption{Profit as a function of $Pc$ and $W_c$, for $P_r=10$}\label{fig:profits_PcWc}
\end{center}
\end{figure}

\begin{figure}%[t]
\begin{center}
\includegraphics[width=\figwidth]{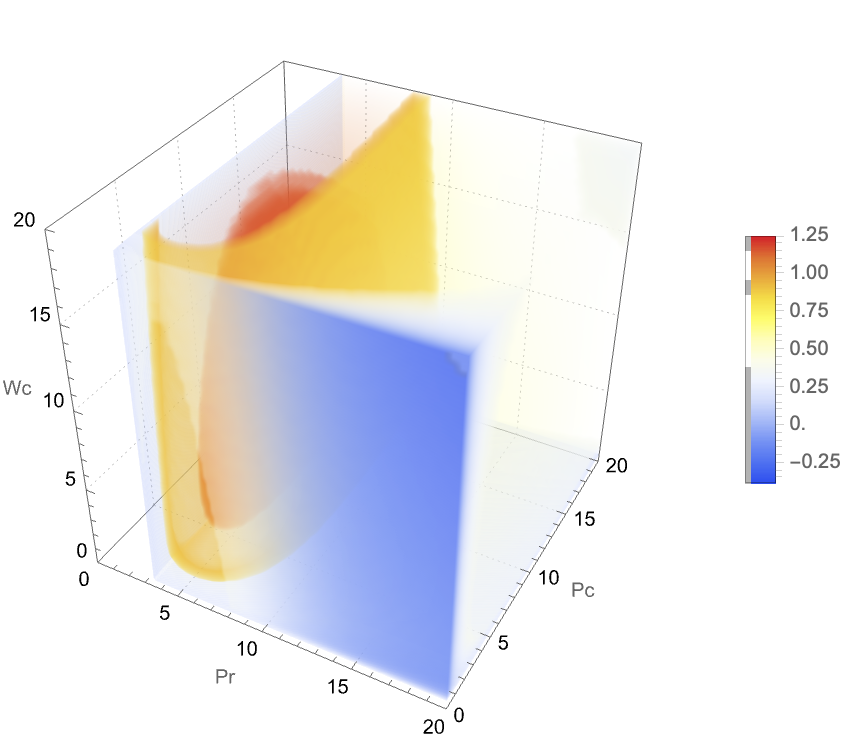}
\caption{Profit as a function of $Pr$, $P_c$ and $W_c$}\label{fig:profits_PrPcWc}
\end{center}
\end{figure}

Profits are plotted in Figs.~\ref{fig:profits_PrPc}, \ref{fig:profits_PcWc} and~\ref{fig:profits_PrPcWc}  as a function of input factors $P_r$, $P_c$ and $W_c$, where parameters are set to the default values of Section~\ref{sec:results}.

Under the usual assumption of profit maximization, the operator will choose input factor quantities $P_r$, $P_c$ and $W_c$ to solve the following problem:
\begin{IEEEeqnarray} {rCCl}\label{eq:PMP}
 \underset{P_r,P_c,W_c}{\max} & & \Pi \\
 \text{subject to } & &  P_r \geq 0 , P_c \geq 0 , W_c \geq 0. \nonumber
\end{IEEEeqnarray} 
Note that Problem~\eqref{eq:PMP} is separable in $P_r$ and $\{ P_c, W_c \}$:
\begin{IEEEeqnarray} {rCCl}\label{eq:PMPc}
 \underset{P_c,W_c}{\max} & & \Pi_c \\
 \text{subject to } & &  P_c \geq 0 , W_c \geq 0. \nonumber
\end{IEEEeqnarray} 
\begin{IEEEeqnarray} {rCCl}\label{eq:PMPr}
 \underset{P_r}{\max} & & \Pi_r \\
 \text{subject to } & &  P_r \geq 0. \nonumber
\end{IEEEeqnarray}
where
\begin{IEEEeqnarray} {rCl}\label{eq:profits_sep}
 \Pi_c & = & p_2 (\rho (P_c, W_c) ) \rho (P_c, W_c) - w_p  P_c - w_w W_c \\
 \Pi_r & = & p_1 (P_r) P_r - w_p P_r .
\end{IEEEeqnarray}

%\section{Analysis}\label{sec:analysis}

\section{Results}\label{sec:results}

We discuss the results in terms of input factor quantities (Fig.~\ref{fig:input_wp},Fig.~\ref{fig:input_ww} and Fig.~\ref{fig:input_alpha}), product quantities produced (Fig.~\ref{fig:product_wp}, Fig.~\ref{fig:product_ww} and Fig.~\ref{fig:product_alpha}), quality perceived (Fig.~\ref{fig:qos_wp}, Fig.~\ref{fig:qos_ww} and Fig.~\ref{fig:qos_alpha}), commodity prices (Fig.~\ref{fig:prices_wp}, Fig.~\ref{fig:qos_ww} and Fig.~\ref{fig:qos_alpha}) and profit (Fig.~\ref{fig:profits_wp}, Fig.~\ref{fig:profits_ww} and  Fig.~\ref{fig:profits_alpha}).

We conduct below comparative statics, that is, we characterize the different optima that result as one parameter is varied across a range of values~\citep{silberberg2000}. Specifically, we conduct three comparative statics. First, we analyze the effect of parameter $w_p$, which is the unit power price; second, the parameter $w_w$, which is the unit bandwidth price; and third, the parameter $\alpha$, which characterizes the willingness to pay for the quality of the sensing functionality.

The optima were obtained by solving numerically the maximization problem~\eqref{eq:PMP}.

The parameters used, if not stated otherwise, are $\gamma=5$, $\gamma_T=1$ and~$\tilde{\gamma}_C=1$ (all three borrowed from \citet{chalise2017b}); and $\alpha=1$, $\beta=1$, $w_p=0.01$ and $w_w=0.01$. 

\subsection{Effect of parameter \texorpdfstring{$w_p$}{wp}}

Here, the parameter $w_p$, the unit price of the transmitted sensing power and communication power, varies between 0.001 and 0.055. Note that this parameter affects both Problem~\eqref{eq:PMPc} and~\eqref{eq:PMPr}.

\begin{figure}%[t]
\begin{center}
\includegraphics[width=\figwidth]{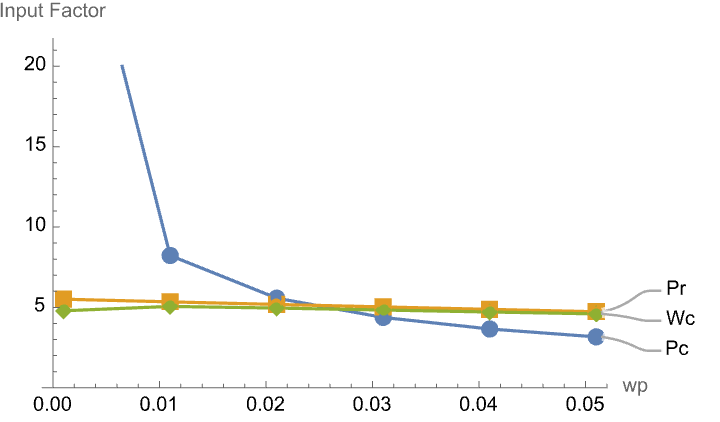}
\caption{$Pr$, $P_c$ and $W_c$ as functions of $w_p$}\label{fig:input_wp}
\end{center}
\end{figure}

\begin{figure}%[t]
\begin{center}
\includegraphics[width=\figwidth]{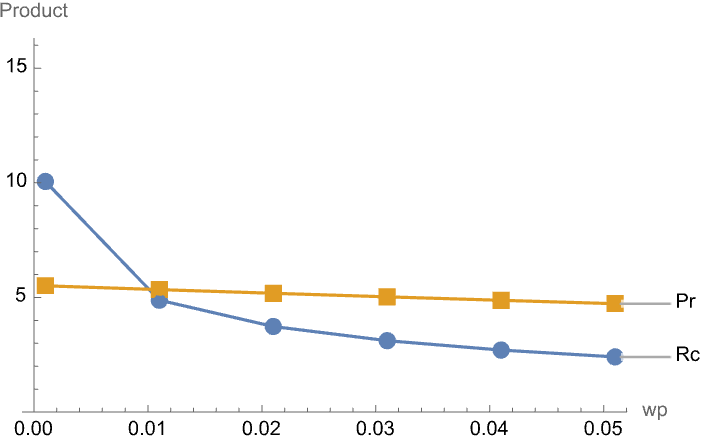}
\caption{$Pr$ and $R_c$ as functions of $w_p$}\label{fig:product_wp}
\end{center}
\end{figure}

\begin{figure}%[t]
\begin{center}
\includegraphics[width=\figwidth]{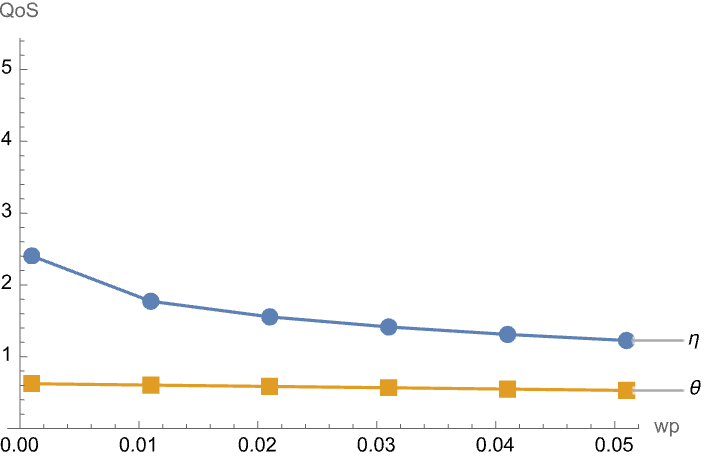}
\caption{Quality metrics as functions of $w_p$}\label{fig:qos_wp}
\end{center}
\end{figure}

\begin{figure}%[t]
\begin{center}
\includegraphics[width=\figwidth]{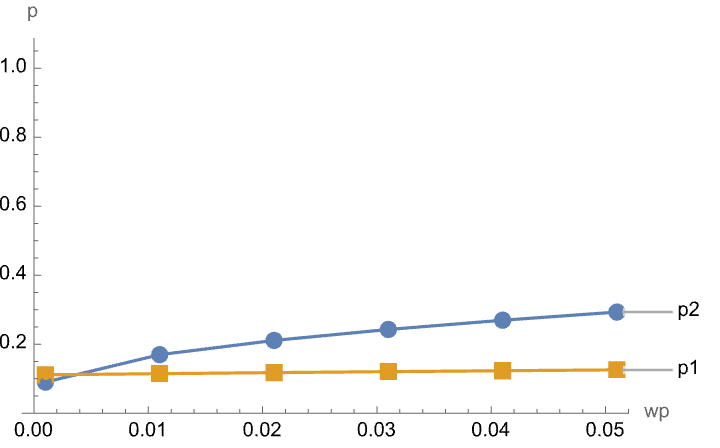}
\caption{Prices as functions of $w_p$}\label{fig:prices_wp}
\end{center}
\end{figure}

\begin{figure}%[t]
\begin{center}
\includegraphics[width=\figwidth]{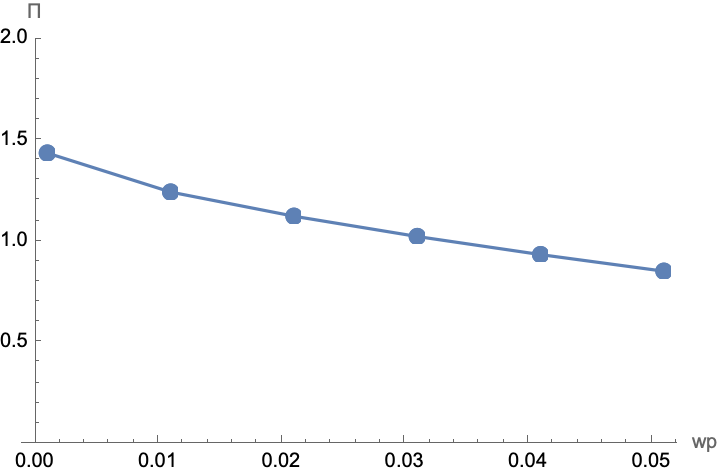}
\caption{Profit as a function of $w_p$}\label{fig:profits_wp}
\end{center}
\end{figure}

The main effect of a $w_p$ increase is, as expected from economic theory, a reduction in the demand of both input factors $P_r$ and $P_c$, the latter being more pronounced (Fig.~\ref{fig:input_wp}); this observation may provide an effective way for the system designer to limit the power consumption, as discussed below in Section~\ref{sec:conclusions}. The reduction in $P_c$ causes a drop in the commodity $R_c$ supplied (Fig.~\ref{fig:product_wp}), and ultimately in the quality of the communication side of the ISAC service $\eta$ (Fig.~\ref{fig:qos_wp}). The reduction in the supply of the commodity $R_c$ is accompanied with a rise in the price $p_2$ (Fig.~\ref{fig:prices_wp}), although the operator profit decreases (Fig.~\ref{fig:profits_wp}).

%\FloatBarrier

\subsection{Effect of parameter \texorpdfstring{$w_w$}{ww}}

Here, the parameter $w_w$, the unit price of the communication bandwidth, varies between 0.001 and 0.055. Note that this parameter only affects Problem~\eqref{eq:PMPc}, so that input factor and commodity $P_r$ and price $p_1$ do not vary.

\begin{figure}%[t]
\begin{center}
\includegraphics[width=\figwidth]{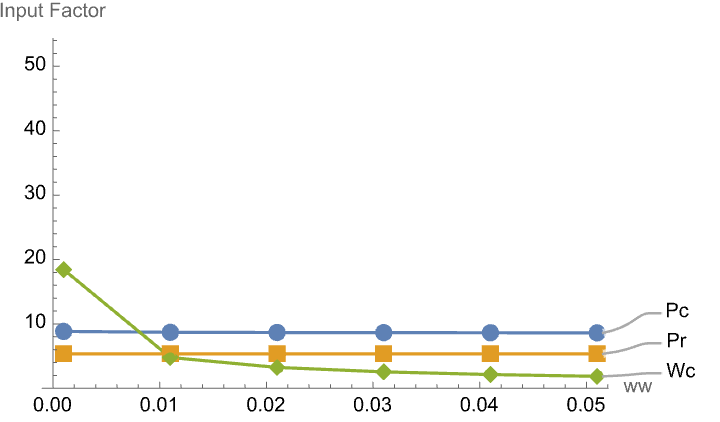}
\caption{$Pr$, $P_c$ and $W_c$ as functions of $w_w$}\label{fig:input_ww}
\end{center}
\end{figure}

\begin{figure}%[t]
\begin{center}
\includegraphics[width=\figwidth]{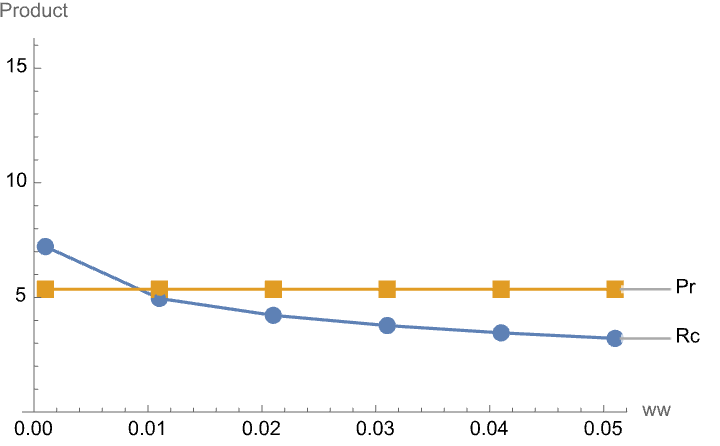}
\caption{$Pr$ and $R_c$ as functions of $w_w$}\label{fig:product_ww}
\end{center}
\end{figure}

\begin{figure}%[t]
\begin{center}
\includegraphics[width=\figwidth]{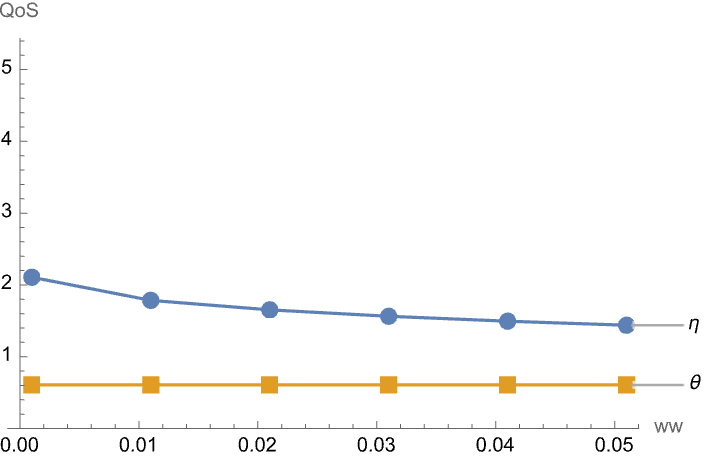}
\caption{Quality metrics as functions of $w_w$}\label{fig:qos_ww}
\end{center}
\end{figure}

\begin{figure}%[t]
\begin{center}
\includegraphics[width=\figwidth]{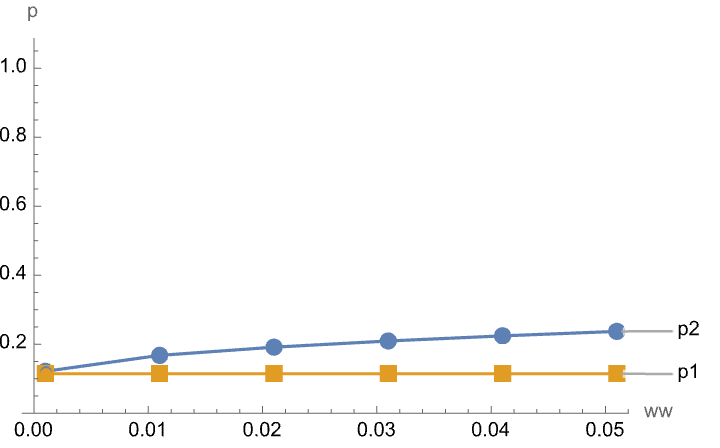}
\caption{Prices as functions of $w_w$}\label{fig:prices_ww}
\end{center}
\end{figure}

\begin{figure}%[t]
\begin{center}
\includegraphics[width=\figwidth]{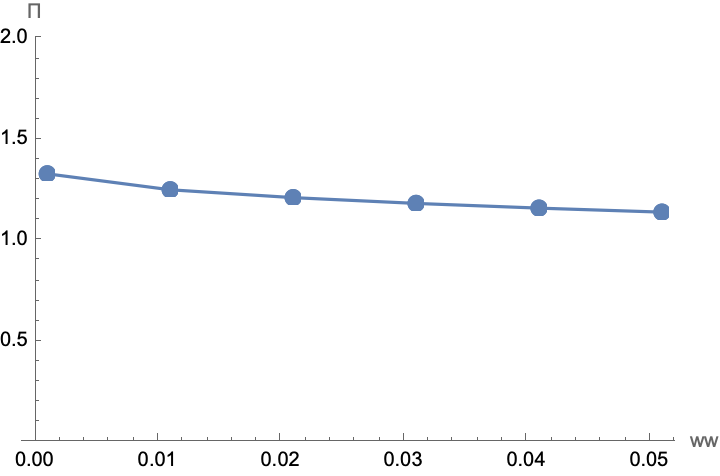}
\caption{Profit as a function of $w_w$}\label{fig:profits_ww}
\end{center}
\end{figure}

The main effect of a $w_w$ increase is, again, a reduction in the demand of input factor $W_c$ (Fig.~\ref{fig:input_ww}); Again, this observation may provide an effective way for the system designer to limit the bandwidth consumption, as it is discussed below in Section~\ref{sec:conclusions}. The reduction in $W_c$ causes a drop in the commodity $R_c$ supplied (Fig.~\ref{fig:product_ww}), and ultimately in the quality of the communication side of the ISAC service $\eta$ (Fig.~\ref{fig:qos_ww}). The reduction in the supply of the commodity $R_c$ is accompanied with a rise in the price $p_2$ (Fig.~\ref{fig:prices_ww}), although again the operator profit decreases (Fig.~\ref{fig:profits_ww}).

%\FloatBarrier

\subsection{Effect of parameter \texorpdfstring{$\alpha$}{alpha}}

Here, the parameter $\alpha$, which is the weight of the sensing part of the ISAC service utility, varies between 0.1 and 2. Note that the weight of the communication part is kept $\beta=1$. And that this parameter only affects Problem~\eqref{eq:PMPr}, so that input factors $P_c$ and $W_c$, commodity $R_c$, price $p_2$ and quality $\eta$ do not vary.

\begin{figure}%[t]
\begin{center}
\includegraphics[width=\figwidth]{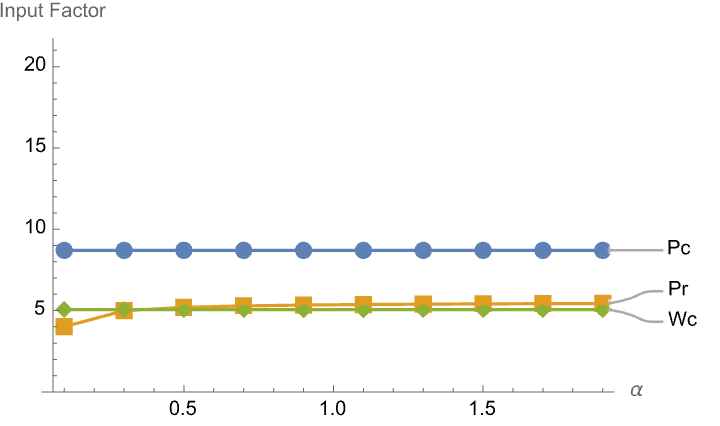}
\caption{$Pr$, $P_c$ and $W_c$ as functions of $\alpha$}\label{fig:input_alpha}
\end{center}
\end{figure}

\begin{figure}%[t]
\begin{center}
\includegraphics[width=\figwidth]{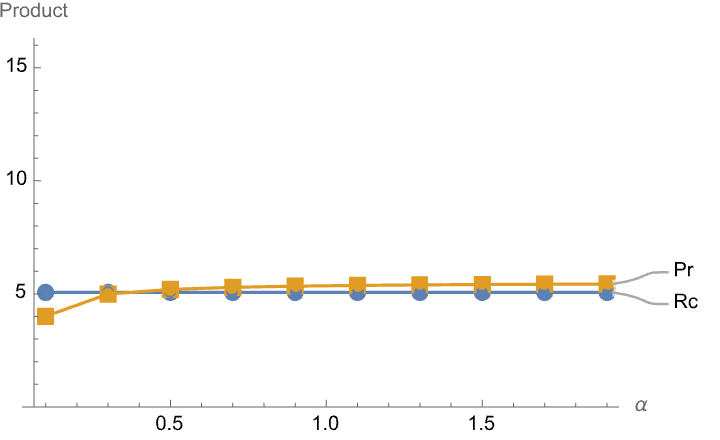}
\caption{$Pr$ and $R_c$ as functions of $\alpha$}\label{fig:product_alpha}
\end{center}
\end{figure}

\begin{figure}%[t]
\begin{center}
\includegraphics[width=\figwidth]{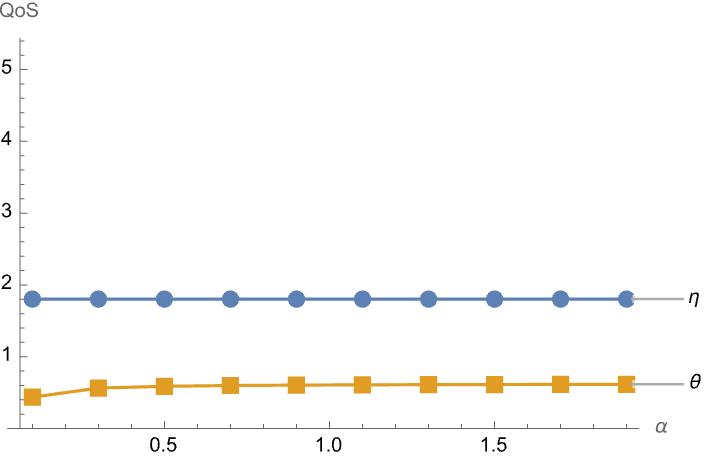}
\caption{Quality metrics as functions of $\alpha$}\label{fig:qos_alpha}
\end{center}
\end{figure}

\begin{figure}%[t]
\begin{center}
\includegraphics[width=\figwidth]{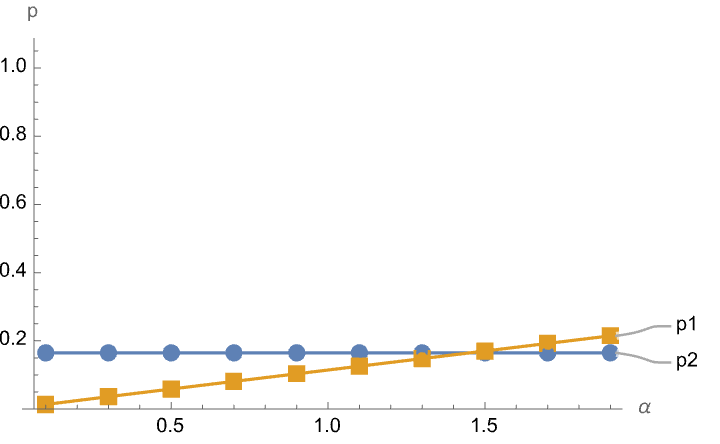}
\caption{Prices as functions of $\alpha$}\label{fig:prices_alpha}
\end{center}
\end{figure}

\begin{figure}%[t]
\begin{center}
\includegraphics[width=\figwidth]{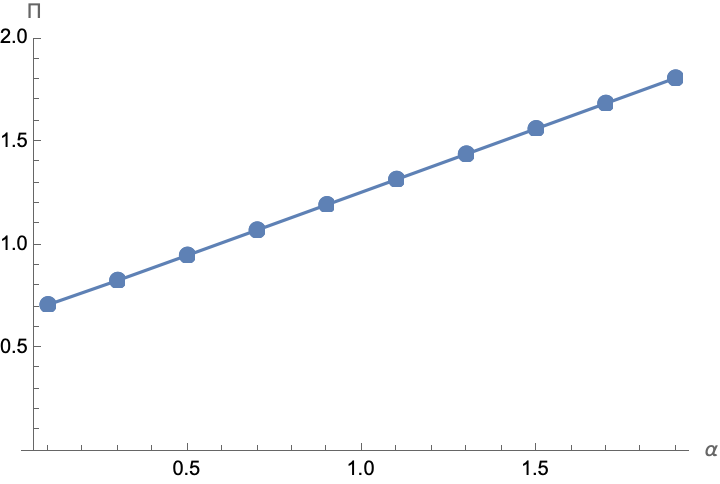}
\caption{Profit as a function of $\alpha$}\label{fig:profits_alpha}
\end{center}
\end{figure}

The effect of an $\alpha$ increase is, as expected from~\eqref{eq:utility}, an increase in the  quality of the sensing part $\theta$ of the ISAC service (Fig.~\ref{fig:qos_alpha}). This drives an increase on the commodity/input factor $P_r$ used for the service (Fig.~\ref{fig:input_alpha} and Fig.~\ref{fig:product_alpha}) and an increase in the unit price $p_1$ ((Fig.~\ref{fig:prices_alpha})). And this drives operator profit upwards (Fig.~\ref{fig:profits_alpha}). Note that the increase in $P_r$ and $\theta$ takes place only for the range of low values of $\alpha$, while the increase in $p_2$ and profit sustains itself for all values of $\alpha$.

%--------------------------------------------------------------
%--------------------------------------------------------------
\section{Conclusions}\label{sec:conclusions}

In this work, we have framed the provision of an ISAC-based service in an economic setting, where the service provision and the resource provisioning are mediated by prices and decisions are taken under utility/profit maximization assumptions.

We have demonstrated that, in a monopolistic scenario, i.e., where one operator provides the ISAC-based service to the market, equilibrium quantities and prices exist. 

From the point of view of an engineering planner, a relevant result has been obtained through comparative statics: prices for power ($w_p$ for $P_r$ and $P_c$) and bandwidth ($w_w$ for $W_c$) resources can be effectively used for controlling their consumption. Specifically, it has been found that this control is more effective on the resources used for producing the communication part of the ISAC service ($P_c$ and $W_c$), while resources for the sensing part ($P_r$) at the equilibrium are less sensitive to the prices. This result provides a more flexible and effective tool for enforcing regulatory limits on both power and bandwidth.

\bibliography{bib}

\end{document}